\documentclass[a4paper]{article}
\usepackage{bm}
\usepackage{graphicx}
\usepackage{amsmath,amssymb}
\newcommand{\erf}{{\rm erf}}
\title{Wave Propagation through an Array of Slits}
\author{Makoto Morinaga, Institute for Laser Science,\\
University of Electro-Communications
\thanks{E-mail address: morinaga@ils.uec.ac.jp}}
\date{}
\begin{document}
\maketitle
\abstract{Propagation of a wave through an array of slits
is theoretically investigated.
The asymptotic expansion of the matrix elements of the propagation operator
is derived and compared with numerical calculations. And then
the eigenmodes and eigenvalues of the propagation operator are
estimated.
Our analysis should provide an insight
into the properties of waveguides composed of opaque masks that
have been proposed recently.}
\section{Introduction}
A new type of waveguides composed of opaque masks (slits or pinholes)
has recently been proposed\cite{dima,MM}. In this waveguide a wave propagates
through a set of identical masks that are aligned on a straight
line with equal spacing (see fig.\ref{sscheme} for the case of slit
array).
The peculiarity of this waveguide is that
no special material is required for its construction
(transparency, high reflectivity, ...).
One possible application of such waveguides
is to integrate light waveguides on a silicon chip.
It can also be used to guide matter waves.
In analyzing the wave propagation in such waveguides,
a continuous model has been used \cite{dima,MM,KO,KO2}.
In this model, the discrete set of opaque masks that constitutes the
waveguide is replaced with a continuous absorbing medium that has a hole
with a cross section identical to the opening
of the masks.
This model predicts that the attenuation per unit length of
the wave propagating in the waveguide is proportional to the square
root of the spacing $L$ between the masks, so that it can be
decreased by reducing $L$.
Though it is known that this model reproduces the experimental results
reasonably,
the physical ground of replacing the discrete set of masks with
a continuous medium is not very clear.
In this paper we consider the wave propagation in an array of
slits and treat the discrete set of slits directly.
% and caluculate the propagation of
%waves through the slits by tracing the diffraction.
We first calculate the asymptotic expansion of the matrix elements
of the propagation operator which is then compared with the
numerical calculations given in the appendix. From these
results we deduce
the eigenmodes and eigenvalues of the propagation operator.

\section{Theoretical model}
\begin{figure}[htbp]
\includegraphics[width=10.5cm]{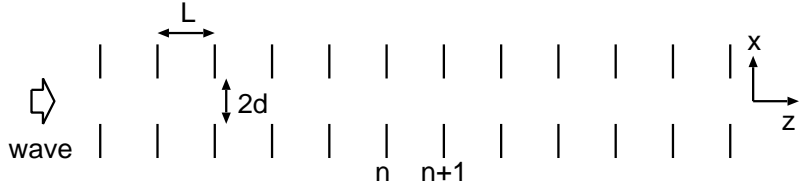}
\caption{Slit array. Identical slits of opening $2d$ (full width)
are aligned on a straight line with spacing $L$ between the slits.
$d\gg\lambda\,(\equiv\frac{2\pi}{k_c})$ is assumed.}
\label{sscheme}
\end{figure}

Suppose a monochromatic wave of a wavenumber $k_c$
is propagating through an array of slits
as depicted in fig.\ref{sscheme}.
We assume that the opening of slits are much larger then
the wavelength of the wave.
We look for the evolution of the transverse wavefunction $\varphi(x)$.
%
%wavenumber: $k_c$\\
%full width of each slit: $2d$\\
%distance between slits: $L$\\
%
The propagation process is divided into two parts:
free propagation $F$ between the two neighbouring slits (for length $L$) and 
masking $M$ of the transverse wavefunction by the slit.
We shall denote the position just after the $n$th slit $P_n$.
Then, the propagation operator $T$ corresponding to the propagation
of the wave from the position $P_n$ to $P_{n+1}$ is written as
$T=MF$.
The masking operator $M$ just masks the transverse wavefunction:
\begin{equation}
 (M\varphi)(x)=
\left\{
\begin{array}{ll}
 \varphi(x)&(|x|<d) \\
 0&(|x|\ge d)
\end{array}
\right. .
\end{equation}
The transverse wavefunction $\varphi(x)$ just after a slit takes
nonzero value only inside the opening of the slit (i.e. $-d<x<d$).
 Such subspace of wavefunctions is spanned by
a set of orthonormal basis functions:
\begin{equation}
 \varphi_m(x)\equiv\left\{
\begin{array}{ll}
 \frac 1{\sqrt d}\sin\frac{(m+1)\pi}{2d}(x+d) & (-d<x<d)\\
 0 & (|x|\ge d)
\end{array}
\right.
\ \ \ (m=0,1,2,...)
\label{basis_f}
\end{equation}
We calculate the matrix elements of $T$ using this basis.
Since $M$ has no effect on this basis functions (i.e.
$M\varphi_m=\varphi_m$), $\langle\varphi_m|T|\varphi_n\rangle
=\langle\varphi_m|F|\varphi_n\rangle$.

\section{Propagation operator $T$}
For a transverse wavefunction that has a transverse wavenumber
$k_x$, the possible values that the longitudinal wavenumber
$k_z$ can take are $\pm\sqrt{k_c^2-k_x^2}$. Here we choose
only $k_z=\sqrt{k_c^2-k_x^2}$,
i.e. we neglect the process of multiple scattering of the wave between slits
in which the wave sometimes propagates backward (i.e. in $-z$ direction).
Thus $F|k_x\rangle=\exp\left(i\sqrt{k_c^2-k_x^2}L\right)|k_x\rangle$.
From the assumption that $d\gg\lambda$, the wavenumber component
of $\varphi_m$ is concentrated in the
region $|k_x|\ll k_c$ so that
$\sqrt{k_c^2-k_x^2}\fallingdotseq
k_c-\frac{k_x^2}{2k_c}$ (paraxial approximation),
and thus $F|k_x\rangle=\exp(
-i\alpha k_x^2)|k_x\rangle$ with $\alpha\equiv\frac L{2k_c}$
(we omit the global phase
factor $\exp(ik_cL)$ hereafter).
The matrix elements $T_{mn}=\langle\varphi_m|T|\varphi_n\rangle$
is calculated as
\begin{equation}
\begin{array}{rl}
 T_{mn}%=&\langle\varphi_m|T|\varphi_n\rangle\\
 =&\langle\varphi_m|F|\varphi_n\rangle\\
 =&\int_{-\infty}^\infty dk_x\,\langle\varphi_m|F|k_x\rangle\langle k_x|\varphi_n\rangle\\
 =&\int_{-\infty}^\infty dk_x\,\langle\varphi_m|k_x\rangle
\exp\left(-i\alpha k_x^2\right)
\langle k_x|\varphi_n\rangle
\end{array}
\label{tmn}
\end{equation}
with
\begin{equation}
 \begin{array}{ll}
 \langle\varphi_m|k_x\rangle&=\int_{-\infty}^\infty
  dx\,\varphi_m(x)\langle x|k_x\rangle\\
  &=\int_{-d}^d dx\,\frac 1{\sqrt d}\sin k_m(x+d)
   \frac 1{\sqrt{2\pi}}e^{ik_xx}\\
%  &=\int_{-d}^d dx\,\frac 1{\sqrt{2\pi d}}\frac 1{2i}
%(e^{ik_m(x+d)}-e^{-ik_m(x+d)})
%   \,e^{ik_xx}\\
%  &=\frac{-1}{2\sqrt{2\pi d}}
%  \left\{
%  \frac{(-1)^{m+1}e^{ik_xd}-e^{-ik_xd}}{k_m-k_x}
%  +\frac{(-1)^{m+1}e^{ik_xd}-e^{-ik_xd}}{k_m+k_x}
%  \right\}\\
%  &=\frac 1{2\sqrt{2\pi d}}
%  \left\{
%  (-1)^{m+1}e^{ik_xd}-e^{-ik_xd}
%  \right\}
%\left(\frac 1{k_x-k_m}-\frac 1{k_x+k_m}\right)\\
  &=\frac{k_m}{\sqrt{2\pi d}}
  \left\{
  (-1)^{m+1}e^{ik_xd}-e^{-ik_xd}
  \right\}
\frac 1{k_x^2-k_m^2}
 \end{array}
\label{iprod}
\end{equation}
where $k_m\equiv\frac{(m+1)\pi}{2d}$.
% We approximate $\exp(-i\alpha k_x^2)$ with $(1+i\alpha k_x^2)^{-1}$.

\section{Matrix elements $T_{mn}$}
\begin{figure}[htbp]
\includegraphics[width=10.5cm]{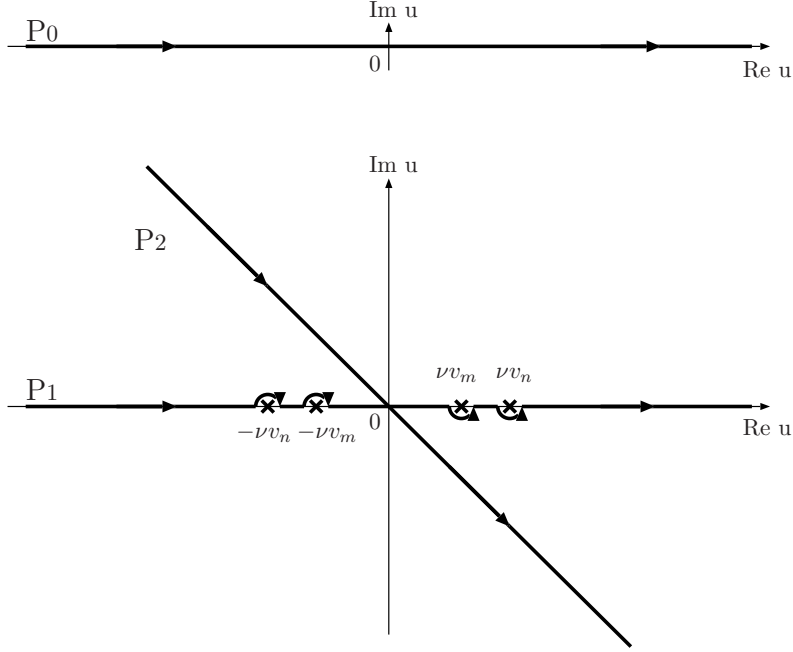}
\caption{Integration path for evaluating $T_{mn}$ (see text).}
%\label{ip}
\label{ip_}
\end{figure}

Now we calculate the asymptotic expansion of the matrix
elements $T_{mn}$ of the propagation operator $T$.
From (\ref{tmn}) and (\ref{iprod}), $T_{mn}$ is calculated as
\begin{equation}
\begin{array}{rl}
 T_{mn}
 =&\frac{k_mk_n}{2\pi d}\int_{-\infty}^\infty dk_x
\frac 1{k_x^2-k_m^2}
\frac 1{k_x^2-k_n^2}\\
  &\ \ \{
  1+(-1)^{m+n}
  +(-1)^me^{2ik_xd}+(-1)^ne^{-2ik_xd}
  \}
  e^{-i\alpha k_x^2}\\
\end{array}
\label{tmnb}
\end{equation}
In the following, we consider only the case when $(-1)^m=(-1)^n$,
because otherwise $T_{mn}=0$.
(\ref{tmnb}) is rewritten using a new variable $u=\sqrt\alpha k_x$
as
\begin{equation}
\begin{array}{rl}
 T_{mn}
 =&\frac{v_mv_n}{2\pi}\int_{-\infty}^\infty \nu^3du
\frac 1{u^2-\nu^2v_m^2}
\frac 1{u^2-\nu^2v_n^2}\\
  &\ \ \times\{
  2%1+(-1)^{m+n}
  +(-1)^me^{2i\frac u\nu}+(-1)^ne^{-2i\frac u\nu}
  \}
  e^{-iu^2}\\
\end{array}
\label{tmnb__}
\end{equation}
with $v_m\equiv k_md=\frac{(m+1)\pi}2$ and
$\nu=\frac{\sqrt\alpha}d=\sqrt{\frac L{2k_cd^2}}
=\frac 1{2\sqrt\pi}\left(\frac{\lambda L}{d^2}\right)^{\frac 12}$.
%
%We are interested in the parameter region when
Hereafter we assume that
$\frac{\lambda L}{d^2}\ll 1$
(and thus $\nu\ll 1$) in which case, as we shall see, the loss of the
propagation is small.
Because the integrand has no pole anywhere, we can change the
integration path from $P_0$ to $P_1$ (see fig.\ref{ip_}),
and then divide the integrand into
3 parts. As a consequence, poles appear at $u=\pm\nu v_m$ and
$u=\pm\nu v_n$.
\begin{equation}
\begin{array}{rl}
 T_{mn}
 =&\frac{v_mv_n}{2\pi}\int_{P_1} \nu^3du
\frac 1{u^2-\nu^2v_m^2}
\frac 1{u^2-\nu^2v_n^2}\\
  &\ \ \times\{
%  \underbrace{1+(-1)^{m+n}}_A
  \underbrace{2}_A
 +\underbrace{(-1)^me^{2i\frac u\nu}}_{B+}
 +\underbrace{(-1)^ne^{-2i\frac u\nu}}_{B-}
  \}
  e^{-iu^2}\\
\end{array}
\label{tmnb___}
\end{equation}
Now we change the integration path from $P_1$ to $P_2$.
%(fig.\ref{ip_}).

\begin{figure}[htbp]
\includegraphics[width=9.5cm]{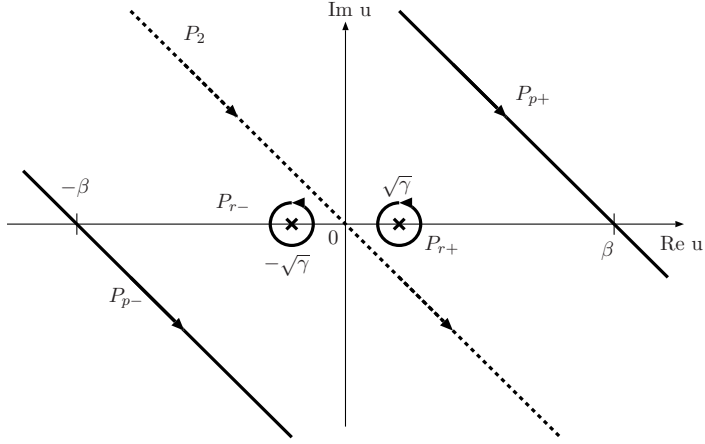}
\caption{Integration path to evaluate $I_1(\nu)$ and
$I_{1\pm}(\nu,\beta)$ (see text).}
\label{ip_ab_}
\end{figure}

\paragraph{The case $m\ne n$}
\begin{equation}
\begin{array}{rl}
 T_{mn}
 =&\frac{v_mv_n}{2\pi(v_m^2-v_n^2)}\int_{P_2} du
\left(\frac\nu{u^2-\nu^2v_m^2}
-\frac\nu{u^2-\nu^2v_n^2}\right)\\
  &\ \ \{
%  \underbrace{1+(-1)^{m+n}}_A
  \underbrace{2}_A
 +\underbrace{(-1)^me^{2i\frac u\nu}}_{B+}
 +\underbrace{(-1)^ne^{-2i\frac u\nu}}_{B-}
  \}
  e^{-iu^2}\\
=&I_A+I_{B+}+I_{B-}
\end{array}
\label{tmnc__}
\end{equation}
By defining
\begin{equation}
 I_1(\gamma)\equiv\int_{P_2}\frac{e^{-iu^2}}{u^2-\gamma}du
\end{equation}
and
\begin{equation}
 I_{1\pm}(\gamma,\beta)\equiv\int_{P_2}
\frac{e^{\pm 2i\beta u}e^{-iu^2}}{u^2-\gamma}du,
\end{equation}
$I_A$ and $I_{B\pm}$ are written as
\begin{equation}
 I_A=\frac{v_mv_n}{\pi(v_m^2-v_n^2)}
 \nu\{I_1(\nu^2v_m^2)-I_1(\nu^2v_n^2)\}
\label{I_A}
\end{equation}
\begin{equation}
 I_{B\pm}=(-1)^m\frac{v_mv_n}{2\pi(v_m^2-v_n^2)}
 \nu\left\{I_{1\pm}\left(\nu^2v_m^2,\frac 1\nu\right)
 -I_{1\pm}\left(\nu^2v_n^2,\frac 1\nu\right)\right\}
\label{I_B}
\end{equation}
%
%\begin{figure}[htbp]
%\includegraphics[width=9.5cm]{ip_gamma.eps}
%\caption{integration path}
%\label{ip_gamma}
%\end{figure}
Taking a variable $s$ as $u=\sqrt{-i}\,s$,
\begin{equation}
\begin{array}{ll}
 I_1(\gamma)&=\sqrt
  i\int_{-\infty}^\infty\frac{e^{-s^2}}{s^2-i\,\gamma}\,ds\\
&=\sqrt ie^{-i\gamma}\int_{-\infty}^\infty
\int_1^{\infty}e^{-(s^2-i\gamma)s'}\,ds'\,ds\\
&=\sqrt{\pi i}\,e^{-i\gamma}
\int_1^{\infty}\frac{e^{i\gamma s'}}{\sqrt{s'}}\,ds'\\
%&=\sqrt{\pi i}\,e^{-i\gamma}\left\{
%\int_0^{\infty}\frac{e^{i\gamma s'}}{\sqrt{s'}}\,ds'
%-\int_0^1\frac{e^{i\gamma s'}}{\sqrt{s'}}\,ds'
%\right\}\\
&=2\sqrt{\pi i}\,e^{-i\gamma}\left\{
\int_0^{\infty}e^{i\gamma s^{\prime\prime 2}}\,ds''
-\int_0^1e^{i\gamma s^{\prime\prime 2}}\,ds''
\right\}\\
&=\sqrt{\pi i}\,e^{-i\gamma}
\sqrt{\frac\pi{-i\gamma}}\left\{1-\erf(\sqrt{-i\gamma})
\right\}\\
%&=e^{-i\gamma}
%\left\{
%\pi i\frac 1{\sqrt\gamma}
%-2\sqrt{\pi i}\sum_{q=0}^{\infty}\frac 1{2q+1}\frac{(i\gamma)^q}{q!}
%\right\}\\
&=e^{-i\gamma}\left\{
\frac{\pi i}{\sqrt\gamma}
-2\sqrt{\pi i}\sum_{n=0}^\infty\frac 1{2n+1}\frac{(i\gamma)^n}{n!}
\right\}\\
&=e^{-i\gamma}
\frac{\pi i}{\sqrt\gamma}
-2\sqrt{\pi i}\sum_{n=0}^\infty\frac{(-2i\gamma)^n}{(2n+1)!!}
%\left\{1+\frac{-2i\gamma}3+\frac{(-2i\gamma)^2}{5!!}
%+\frac{(-2i\gamma)^3}{7!!}+...\right\}
\end{array}
\label{I1}
\end{equation}
where $\erf$ is the error function defined by
\begin{equation}
 \erf(z)\equiv\frac 2{\sqrt\pi}\int_0^ze^{-t^2}dt
% =\frac 1{\sqrt\pi}\int_0^{z^2}\frac{e^{-u}}{\sqrt u}du
 =\frac 2{\sqrt\pi}\left(z-\frac{z^3}3+\frac{z^5}{10}
 -\frac{z^7}{42}+...\right).
\end{equation}
The formula used to derive the series expansion of the last line
of (\ref{I1}) is presented in the appendix\ref{formula}.
To evaluate $I_{1\pm}(\gamma,\beta)$, the integration path is
changed from $P_2$ to $P_{r\pm}+P_{p\pm}$ (fig.\ref{ip_ab_}):
\begin{equation}
 I_{1\pm}(\gamma,\beta)=\int_{P_{r\pm}+P_{p\pm}}
\frac{e^{\pm 2i\beta u}e^{-iu^2}}{u^2-\gamma}du
=I_{1r\pm}(\gamma,\beta)+I_{1p\pm}(\gamma,\beta)
\end{equation}
$I_{1r\pm}(\gamma,\beta)$ is calculated by evaluating the residue
of the integrand at $u=\pm\sqrt\gamma$ giving
\begin{equation}
I_{1r\pm}(\gamma,\beta)
% \int_{P_{r\pm}}
%\frac{e^{\pm 2i\beta z}e^{-iz^2}}{z^2-\gamma}dz
%=\pm 2\pi i\frac{e^{2i\beta\sqrt\gamma}e^{-i\gamma}}{\pm 2\sqrt\gamma}
=\pi i\frac{e^{2i\beta\sqrt\gamma}e^{-i\gamma}}{\sqrt\gamma}
\end{equation}
To evaluate $I_{1p\pm}(\gamma,\beta)$, we take a variable $s$
as $u=\sqrt{-i}\,s\pm\beta$ so that
\begin{equation}
 I_{1p\pm}(\gamma,\beta)=\sqrt i\,e^{i\beta^2}\int_{-\infty}^{\infty}
\frac{e^{-s^2}}{(s\pm\sqrt i\beta)^2-\gamma}ds
\end{equation}
We now assume that $|\gamma|\ll 1$ and $|\beta|\gg 1$
(because $I_{1\pm}$ is used in the form (\ref{I_B})
and $\nu\ll 1$)
so that $I_{1p\pm}$ can be expanded as
\begin{equation}
 I_{1p\pm}(\gamma,\beta)=\sqrt i\,e^{i\beta^2}
\sum_{n=0}^\infty g_{n\pm}(\beta)\,\gamma^n
%\left(-\frac 1{\beta^2}+\frac\gamma{\beta^4}+...\right)
\end{equation}
with
\begin{equation}
 g_{n\pm}(\beta)\equiv\int_{-\infty}^\infty
 \frac{e^{-s^2}}{(s\pm\sqrt i\beta)^{2(n+1)}}\,ds
 =\frac{(-i)^{n+1}\sqrt\pi}{\beta^{2(n+1)}}+...
\end{equation}
Terms that appear in the right-hand sides of (\ref{I_A})
and (\ref{I_B}) are evaluated as
\begin{equation}
 I_1(\nu^2v_m^2)=e^{-i\nu^2v_m^2}
 \frac{\pi\,i}{\nu v_m}-2\sqrt{\pi i}
\sum_{n=0}^\infty\frac{(-2i\,\nu^2v_m^2)^n}{(2n+1)!!}
\end{equation}
\begin{equation}
 I_{1r\pm}\left(\nu^2v_m^2,\frac 1\nu\right)
=\pi i\frac{(-1)^{m+1}e^{-i\nu^2v_m^2}}{\nu v_m}
\end{equation}
Now we are ready to evaluate $T_{mn}$. The contribution of the
first term of $I_1$ cancels with the contribution of $I_{1r\pm}$.
\begin{equation}
\begin{array}{ll}
 T_{mn}&=\frac{v_mv_n}{\pi}\nu
(-2\sqrt{\pi i})
\sum_{n=1}^\infty\frac{(-2i\,\nu^2)^n}{(2n+1)!!}
\frac{v_m^{2n}-v_n^{2n}}{v_m^2-v_n^2}+T_{osc}
\\
&=-4\sqrt{\frac{-i}\pi}v_mv_n\nu^3
\sum_{n=1}^\infty\frac{(-2i\,\nu^2)^{n-1}}{(2n+1)!!}
\frac{v_m^{2n}-v_n^{2n}}{v_m^2-v_n^2}+T_{osc}
\\
&=-\frac 43\sqrt{\frac{-i}\pi}v_mv_n\nu^3
\left(
1+\frac{-2i\nu^2}5\frac{v_m^{4}-v_n^{4}}{v_m^2-v_n^2}
+\frac{(-2i\nu^2)^2}{35}\frac{v_m^{6}-v_n^{6}}{v_m^2-v_n^2}
+...\right)+T_{osc}
\end{array}
\end{equation}
The term $T_{osc}$ is the contribution from $I_{p\pm}$ and
oscillates rapidly % with $\nu$
as $e^{i\beta^2}=e^{i\nu^{-2}}$ when
$\nu\to 0$.
Its leading term is proportional to $\nu^7$ in magnitude:
\begin{equation}
 T_{osc}=\sqrt{\frac{-i}\pi}v_mv_ne^{i\nu^{-2}}\nu^7
 \left\{1+O(\nu)
\right\}
\label{T_osc}
\end{equation}

\paragraph{The case $m=n$}
\begin{equation}
\begin{array}{rl}
 T_{mm}
 =&\frac{v_m^2}{2\pi}\int_{-\infty}^\infty \nu^3du
\left(\frac 1{u^2-\nu^2v_m^2}\right)^2
 \{
  \underbrace{2}_A
 +\underbrace{(-1)^me^{2i\frac u\nu}}_{B+}
 +\underbrace{(-1)^me^{-2i\frac u\nu}}_{B-}
  \}
  e^{-iu^2}\\
=&I_A+I_{B+}+I_{B-}
\end{array}
\label{tmmb__}
\end{equation}

By defining
\begin{equation}
 I_2(\gamma)\equiv\int_{P_2}\frac{e^{-iu^2}}{(u^2-\gamma)^2}du
\end{equation}
and
\begin{equation}
 I_{2\pm}(\gamma,\beta)\equiv
\int_{P_2}
\frac{e^{\pm 2i\beta u}e^{-iu^2}}{(u^2-\gamma)^2}du,
%=\int_{P_{r\pm}+P_{p\pm}}
%\frac{e^{\pm 2i\beta u}e^{-iu^2}}{(u^2-\gamma)^2}du,
\end{equation}
$I_A$ and $I_{B\pm}$ are written as
\begin{equation}
 I_A=\frac{v_m^2}{\pi}\,\nu^3I_2(\nu^2v_m^2)
\end{equation}
and
\begin{equation}
 I_{B\pm}=(-1)^m\frac{v_m^2}{2\pi}\,\nu^3I_{2\pm}
 \left(\nu^2v_m^2,\frac 1\nu\right)
\end{equation}
where $I_2$ and $I_{2\pm}$ are evaluated from $I_1$ and $I_{1\pm}$
as follows:
\begin{subequations}
\begin{eqnarray}
 I_2(\gamma)=I_1'(\gamma)=&
\partial_\gamma\left(
e^{-i\gamma}\frac{\pi i}{\sqrt\gamma}\right)
-4\sqrt{-\pi i}\,e^{-i\gamma}\sum_{n=0}^\infty
\frac 1{(2n+1)(2n+3)}\frac{(i\gamma)^n}{n!}\\
=&\partial_\gamma\left(
e^{-i\gamma}\frac{\pi i}{\sqrt\gamma}\right)
-4\sqrt{-\pi i}\sum_{n=0}^\infty
(n+1)\frac{(-2i\gamma)^n}{(2n+3)!!}
\end{eqnarray}
\end{subequations}
\begin{equation}
 I_{2\pm}(\gamma,\beta)=\partial_\gamma I_{1\pm}(\gamma,\beta)
 =I_{2r\pm}(\gamma,\beta)+I_{2p\pm}(\gamma,\beta)
\end{equation}
Here $I_{2r\pm}$ and $I_{2p\pm}$ are defined as
\begin{eqnarray}
  I_{2r\pm}(\gamma,\beta)\equiv&\partial_\gamma I_{1r\pm}(\gamma,\beta)
 =&e^{2i\beta\sqrt\gamma}
  \partial_\gamma\left(e^{-i\gamma}\frac{\pi i}{\sqrt\gamma}\right)
  -\frac{\pi\beta}{\gamma}e^{2i\beta\sqrt\gamma}
   e^{-i\gamma}
\\
  I_{2p\pm}(\gamma,\beta)\equiv&\partial_\gamma I_{1p\pm}(\gamma,\beta)
  =&\sqrt{-\pi i}e^{i\beta}\frac 1{\beta^4}+...
\end{eqnarray}
Finally $T_{mm}$ is calculated as
\begin{subequations}
\begin{eqnarray}
 T_{mm}=&e^{-i\nu^2v_m^2}\left\{1-4\sqrt{\frac{-i}\pi}\nu^3v_m^2
\sum_{n=0}^\infty
\frac 1{(2n+1)(2n+3)}\frac{(i\nu^2v_m^2)^n}{n!}\right\}
+T_{osc}\\
 =&e^{-i\nu^2v_m^2}-4\sqrt{\frac{-i}\pi}\nu^3v_m^2
\sum_{n=0}^\infty
(n+1)\frac{(-2i\nu^2v_m^2)^n}{(2n+3)!!}
+T_{osc}
\end{eqnarray}
\end{subequations}
where the leading term of
the oscillating term $T_{osc}$ is same as in (\ref{T_osc}) with $n=m$.
$T_{mn}$ up to the order of $O(\nu^4)$ can be summarized as below
for both $m=n$ and $m\ne n$ cases:
\begin{subequations}
\begin{eqnarray}
  T_{mn}=&
\delta_{mn}\,e^{-i\nu^2v_m^2}-\frac{1+(-1)^{m+n}}2
\frac 43\sqrt{\frac{-i}\pi}v_mv_n\nu^3
+O(\nu^5)\\
=&\delta_{mn}\,e^{-i\frac{(m+1)^2\pi}{16}\left(\frac{\lambda L}{d^2}\right)}
-\frac{1+(-1)^{m+n}}2\frac{(m+1)(n+1)(1-i)}{24\sqrt 2}
 \left(\frac{\lambda L}{d^2}\right)^{\frac 32}
+O\left(\left(\frac{\lambda L}{d^2}\right)^{\frac 52}\right)
\label{tmn_b}
\end{eqnarray}
\end{subequations}

\section{Eigenmodes}
Transverse wavefunctions of eigenmodes of a slit waveguide
at the position just after a slit is given as eigenstates of the evolution
operator $T$.
According to the perturbation theory, $m\,$th eigenstate is calculated
as
\begin{equation}
 |\tilde\varphi_m\rangle=
 |\varphi_m\rangle
 +\sum_{n\ne m}\frac{T_{nm}}{T_{mm}-T_{nn}}|\varphi_n\rangle
\label{petur}
\end{equation}
From (\ref{tmn_b}) we find for $n\ne m$ and small
$\frac{\lambda L}{d^2}$
\begin{equation}
 |T_{mn}|\sim\frac{(m+1)(n+1)}{24}
 \left(\frac{\lambda L}{d^2}\right)^{\frac 32}
%+O\left(\left(\frac{\lambda L}{d^2}\right)^{\frac 52}\right)
\label{Tmn}
\end{equation}
\begin{equation}
 |T_{mm}-T_{nn}|
\sim\left|\frac{(m+1)^2-(n+1)^2}{16}\right|
 \pi\,\frac{\lambda L}{d^2}
\label{DT}
\end{equation}
so that $T_{nm}$ approaches faster to 0 than $T_{mm}-T_{nn}$
when $\frac{\lambda L}{d^2}\rightarrow 0$.
We also see that the net contribution of states other than $|\varphi_m\rangle$
vanishes in (\ref{petur}) because
\begin{equation}
 \sum_{n\ne m}
 \left|\frac{T_{nm}}{T_{mm}-T_{nn}}\right|^2\rightarrow 0
\end{equation}
when $\frac{\lambda L}{d^2}\rightarrow 0$.
In summary, for small $\frac{\lambda L}{d^2}$, $|\varphi_m\rangle$
can be regarded as the $m\,$th eigenmodes of the slit waveguide
and $T_{mm}$ is the eigenvalue for that mode.
From the numerical calculations given in the appendix
(see fig.\ref{diag} and fig.\ref{odiag}), we see that
such approximation is valid roughly for $\frac{\lambda L}{d^2}
\lesssim 0.3$.
The amplitude attenuation per one slit (i.e. for length $L$)
for example, is given by
\begin{equation}
 1-|T_{mm}|\sim
 \frac{(m+1)^2}{24\sqrt 2}
 \left(\frac{\lambda L}{d^2}\right)^{\frac 32}.
\label{1-Tmm}
\end{equation}
The corresponding value that the continuous model predicts is\cite{MM}
\begin{equation}
 1-|T_{mm}|\sim
 \frac{(m+1)^2\sqrt{\pi}}{16\sqrt\xi}
 \left(\frac{\lambda L}{d^2}\right)^{\frac 32},
\label{1-Tmmc}
\end{equation}
where $\xi$ is a dimensionless parameter related to the
absorbance of the continuous medium that can not be determined
by the continuous model itself.
We note that (\ref{1-Tmmc}) becomes identical
to (\ref{1-Tmm}) by putting $\xi=\frac 92\pi$. 
%which reproduces the same dependence on $\frac{\lambda L}{d^2}$
%and $m$ predicted by the continuous model\cite{MM}.
%
\section{Conclusion}
In this paper we determined the transverse mode functions and their
attenuation rates of a slit waveguide by calculating the matrix
elements of the propagation operator and then evaluating its
eigenmodes and eigenvalues.
Our calculation not only confirms the power law of the
attenuation rate on the
waveguide parameters predicted by the continuous model, but also gives
the absolute value of the attenuation rate which could not be obtained
by the continuous model.

\section*{Acknowledgements}
This work was partly supported by the Photon Frontier Network Program
(MEXT).

\appendix
\section{Numerical calculations}
\begin{figure}[htbp]
\includegraphics[width=12.5cm]{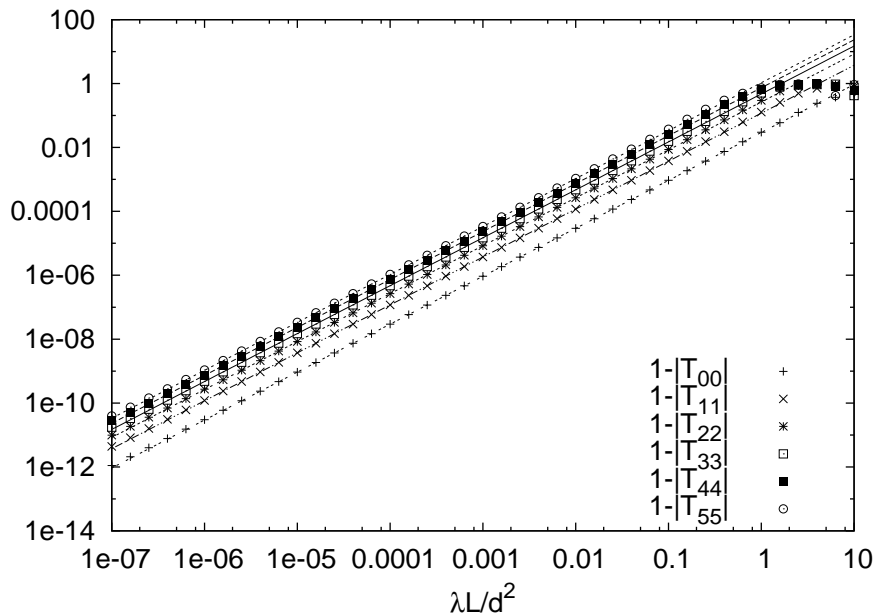}
\caption{Plot of $1-|T_{mm}|$ (amplitude attenuation of the
$m\,$th mode per length $L$). Lines are the corresponding asymptotic values
calculated using (\ref{1-Tmm}).}
\label{diag}
\end{figure}

\begin{figure}[htbp]
\includegraphics[width=12.5cm]{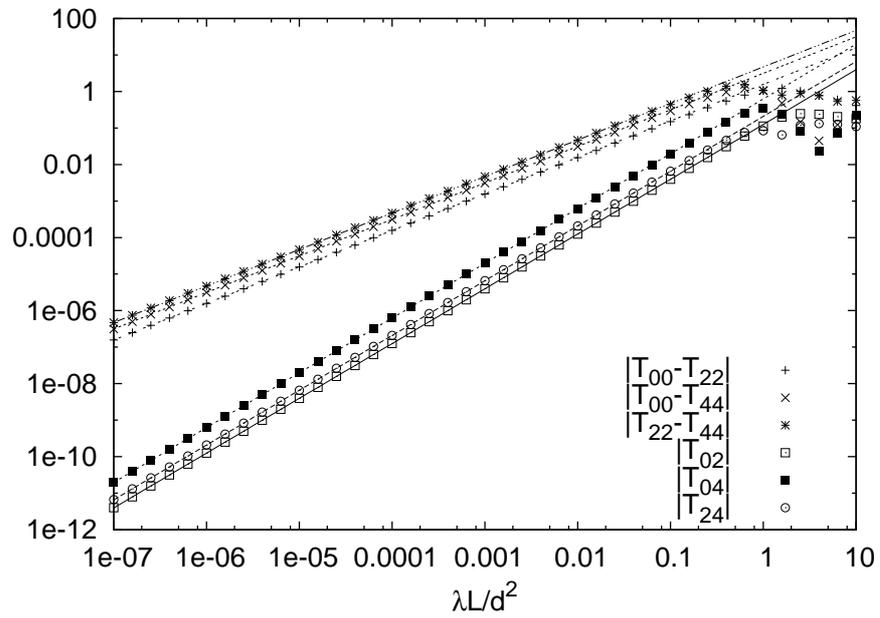}
\caption{Magnitude of the off diagonal elements
($|T_{02}|$, $|T_{04}|$, and $|T_{24}|$) v.s. the magnitude
of difference of the diagonal elements ($|T_{00}-T_{22}|$,
$|T_{00}-T_{44}|$, and $|T_{22}-T_{44}|$) of the propagation
operator $T$.
Lines are the corresponding asymptotic values
calculated using (\ref{Tmn}) and (\ref{DT}).}
\label{odiag}
\end{figure}

We evaluate the matrix elements $T_{mn}$ of the propagation operator
$T$ numerically and compare with the calculation presented in the main part
of this paper.
First we prepare the basis functions $\varphi_m(x)$
(defined in (\ref{basis_f})) in a
discrete space consists of $2^{17}=131072$ sampling points of which
an interval consists of 25601 points corresponds to the
opening of the slit. Then we calculate $\langle k|\varphi_m\rangle$
by taking FFT of $\varphi_m(x)$ and finally evaluate $T_{mn}$
using the expression (\ref{tmn}).
In fig.\ref{diag} we plot $1-|T_{mm}|$ with $m=0,1,2,3,4,5$ where
as in fig.\ref{odiag} differences of the diagonal elements
($|T_{00}-T_{22}|$, $|T_{00}-T_{44}|$, and $|T_{22}-T_{44}|$)
are plotted along with the off diagonal elements
($|T_{02}|$, $|T_{04}|$, and $|T_{24}|$).
In both figures, the corresponding values calculated using the
asymptotic expressions ((\ref{1-Tmm}), (\ref{Tmn}), and (\ref{DT}))
are also plotted with lines.
%Horizontal axes in both figures are $\frac{\lambda L}{d^2}$.
From fig.\ref{odiag}, we see that the magnitude of off diagonal
terms relative to the differences of the diagonal elements decreases
with decreasing $\frac{\lambda L}{d^2}$. This means that the state
$\varphi_m$ approaches to the eigenstate of $T$ when
$\frac{\lambda L}{d^2}\rightarrow 0$
(practically $\varphi_m$ can be regarded as good eigenstate for
$\frac{\lambda L}{d^2}\lesssim 0.1$).
$T_{mm}$ becomes eigenvalue of $T$ for small $\frac{\lambda L}{d^2}$
and then $1-|T_{mm}|$ has the meaning of amplitude attenuation per
one slit (i.e. per length $L$).
%In fig.\ref{diag}, $1-|T_{mm}|$ is plotted along with
%the asymptotic value given by (\ref{1-Tmm}).

\section{Useful formula}
\label{formula}
We give the series expansion of a function $g(\gamma)$ (defined below)
which is used to derive the expansion of the last line of (\ref{I1}).
\[
 g(\gamma)\equiv e^{-i\gamma}\int_0^1e^{i\gamma s^2}ds
 =\int_0^1e^{-i\gamma(1-s^2)}ds
 =\sum_{n=0}^\infty\frac{(-i\gamma)^n}{n!}f_n
\]
where $f_n\equiv\int_0^1(1-s^2)^n\,ds$
(and thus $f_0=1$).
By integrating the following expression from s=0 to 1
\[
 [s(1-s^2)^n]'=(1-s^2)^n-2ns^2(1-s^2)^{n-1}=(2n+1)(1-s^2)^n-2n(1-s^2)^{n-1}
\]
we obtain
\[
 f_n=\frac{2n}{2n+1}f_{n-1}=\frac{2^nn!}{(2n+1)!!}
\]
so that
\[
 g(\gamma)=\sum_{n=0}^\infty\frac{(-2i\gamma)^n}{(2n+1)!!}.
\]

\end{document}